\documentclass[useAMS,usenatbib]{mn2e}
\usepackage{graphicx}
\usepackage{amsmath}

\title[Effect of energy-gain variance on shock-acceleration]{The effect of energy amplification variance on the shock-acceleration}
\author[Junichi Aoi, Kohta Murase and Shigehiro Nagataki]
{Junichi Aoi$^{}$\thanks{E-mail:aoi@yukawa.kyoto-u.ac.jp}, Kohta Murase$^{}$
and Shigehiro Nagataki$^{}$\\
$^{}$Yukawa Institute for Theoretical Physics,Kyoto University, Oiwake-cho,Kitashirakawa, Sakyo-ku, Kyoto, 606-8502, Japan\\}

\begin{document}

\date{Accepted 2007 October 19. Received 2007 October 12; in original form 2007 May 25}

\pagerange{\pageref{firstpage}--\pageref{lastpage}} \pubyear{}

\maketitle

\label{firstpage}

\begin{abstract}
The shock-acceleration theory predicts a power-law energy spectrum in the test particle approximation, and 
there are two ways to calculate a power-law index, Peacock's approximation and Vietri's formulation. 
In Peacock's approximation, it is assumed that particles cross a shock front many times and energy-gains 
for each step are fully uncorrelated. On the other hand, correlation of the distribution of an energy-gain 
factor for a particle is considered in Vietri's formulation. We examine how Peacock's approximation differs 
from Vietri's formulation. It is useful to know when we can use Peacock's approximation because Peacock's 
approximation is simple to derive the power-law index. In addition, we focus on how the variance of the 
energy-gain factor has an influence on the difference between Vietri's formulation and Peacock's approximation. 
The effect of the variance has not been examined well until now. For demonstration, we consider two cases 
for the scattering in the upstream: the large-angle scattering (model A) and the regular deflection by 
large-scale magnetic fields (model B). Especially there is no correlation among the distribution of an 
energy-gain factor for every step in model A. In this model, we see the power-law index derived from Peacock's 
approximation differs from the one derived from Vietri's formulation when we consider the mildly-relativistic 
shock, and the variance of the energy-gain factor affects this difference. We can use Peacock's approximation 
for a non-relativistic shock and a highly-relativistic shock because the effect of the variance is hidden. 
In model B, we see the difference of the power-law converging along the shock velocity. 

\end{abstract}

\begin{keywords}
acceleration of particles - shock waves - methods:analytical - cosmic rays.

\end{keywords}

\section{Introduction}

Galactic cosmic rays with energies $E \le 10^{15}$ eV, below the so called 'knee' in the cosmic ray spectrum, 
are thought to originate from shocks of supernova remnants. The shock-acceleration theory was proposed by 
various authors independently to explain the origin of Galactic cosmic rays. \cite{b2}, \citet{b14}, \citet{b4} 
and \citet{b5} studied the shock-acceleration for non-relativistic shocks. Their theory relies on multiple 
scattering of charged particles which results in round-trips between the upstream and the downstream of a shock. 
In the test particle approximation, shock-acceleration theory predicts that a shock produces energetic particles 
whose energy spectrum obeys a power-law which explains the observation of the high energy cosmic rays well. 
A power-law is obtained by two ways. 1) Solving the convection-diffusion transport equation with appropriate 
boundary conditions. 2) Considering an interaction of test particles with a shock front from the kinetic 
theory viewpoint. These independent two theories predict the same result in spite of different approaches. 

Cosmic rays with energies $E \sim 10^{18.5}-10^{20.5}$ eV are generally believed to come from extragalactic 
origins, based on their harder spectrum, isotropic arrival directions on the sky and the fact that they 
are not confined by the Galactic magnetic fields. Some of authors consider cosmic ray production for the 
relativistic shocks (\citet{b21}; \citet{b19}; \citet{b15}). The relativistic shock-acceleration is first 
considered by \citet{b17} considering relativistic kinematics based on the Bell's theory taking account of 
the beaming effect. Later, \citet{b12},\citet{b13} solved the transport equation with appropriate collision 
operators. They focussed on the diffusion approximation. 

On the other hand, \citet{b9} and \citet{b20} formulated the shock-acceleration for the arbitrary shock speeds without 
the diffusion approximation. Their theories include the non-diffusive effect which is important for relativistic 
shocks and reproduce the same result in the non-relativistic limit. Both of them are constructed in two parts. 
First, they calculated $ P_u(\mu_0,\mu)$ ($ P_d(\mu_0,\mu)$ ) which denotes the conditional probability that 
a particle entering into the upstream (downstream) along a direction $\mu_0$ will leave toward the downstream 
(upstream) along a direction $\mu$. They used some phenominological scattering models to calculate $P_u$ and 
$P_d$. Second, they calculated a power-law index using different ways. \citet{b17} calculated a power-law index 
assuming that particles cross a shock front many times and an energy-gain of each step is fully uncorrelated 
(Peacock's approximation). \citet{b9} used Peacock's approximation. On the other hand, \citet{b20} solved the 
transport equation exactly and calculated a power-law index (Vietri's formulation). He considered correlation 
among energy-gains of a particle. He used his original formulation to calculate a power-law index. Vietri's 
formulation gives the different result from Peacock's approximation in general but both approaches give the 
same result in the Newtonian limit (see also \citet{b6}).

To calculate a power-law index, we have to calculate $P_u$ and $P_d$ using some scattering models. \citet{b9} 
considered the large-angle scattering in both the upstream and the downstream to calculate $P_u$ and $P_d$ 
(Model A). The large-angle scattering model mimics the scattering in strongly turbulent fields. It is often 
supposed that the turbulence can be strong in astrophysical shock environments (see \citet{b7}, and references 
therein). In this model, the initial information of the distribution function is lost when particles are 
scattered. So an energy-gain factor of each step is uncorrelated. Moreover, \citet{b8} pointed out necessity 
to consider the regular deflection as far as relativistic shocks are concerned (Model B). It was understood that 
return of the particles to a shock surface from the upstream region can be warranted even in the absence of 
scattering, provided background magnetic fields are at an angle with the shock normal. This is due to the fact 
particles return to the shock surface from the upstream by the regular deflection before scattering occurs. 

In this paper, we examine how the variance of the energy-gain factor influences the difference between Vietri's 
formulation and Peacock's approximation. It is useful to know when we can use Peacock's approximation because 
Peacock's approximation is simple to derive a power-law index. It is also important to understand Vietri's 
formulation and Peacock's approximation in detail because we can examine validity of the works in which 
Peacock's approximation is used. First, we calculate $P_u$ and $P_d$ in model A (where the large-angle 
scattering occurs at both sides of a shock) which is suitable for examining the effect of the variance of 
the energy-gain factor because there is no effect of the correlation in this model. Then we calculate a power-law 
index using both Peacock's approximation and Vietri's formulation. Next, we consider model B where the 
large-angle scattering occurs in the downstream and a particle is deflected by large-scale magnetic fields in 
the upstream. We calculate probability functions and a power-law index using Peacock's approximation and 
Vietri's formulation in the same way as the case of model A. Finally, we examine how the power-law index derived 
from Peacock's approximation differs from the one derived from Vietri's formulation to examine the effect of the 
variance. In model A, we see that the variance affects the difference between Peacock's approximation and Vietri's 
formulation, and explain what the effect of the variance means physically. To examine the above fact, we show the 
power-law index changing the shock velocity from a non-relativistic one to a highly-relativistic one. 
\citet{b16} calculated a power-law index with $0.04\leq\Gamma_s\beta_s\leq10$ considering the regular deflection, 
where $\beta_s$ is the shock velocity and $\Gamma_s$ is the Lorentz factor of the shock velocity . We extend this 
calculation to the highly-relativistic range. 

The plan of this paper is as follows. In section 2 we briefly summarize the theoretical framework introduced in 
\citet{b9}. We also review Peacock's approximation and Vietri's formulation. In section 3, first, we consider 
the large angle scattering in both the upstream and the downstream. Next, we consider the regular deflection 
in the upstream and the large angle scattering in the downstream. We also discuss what makes Peacock's 
approximation inadequate. Discussion and conclusion are presented in section 4. 

\section{Method of Calculation}
In this paper we use the shock-acceleration formulation which is applicable to any value of the shock speed in the 
static situation and calculate a power-law index using Peacock'approximation and Vietri's formulation. We assume the 
test particle approximation and adopt the large-angle scattering. We also consider the regular deflection by 
large-scale magnetic fields. (we take the unit $c=1$)
\subsection{Shock structure}
First, we have to determine a shock structure by solving jump conditions and an equation of state. (e.g., 
\citet{b11}) Relativistic jump conditions are written as
\begin{equation}
\Gamma_u\beta_un_u\quad = \Gamma_d\beta_dn_d,
\end{equation}
\begin{equation}
\Gamma^2_u\beta_u(\epsilon_u+p_u) = \Gamma^2_d\beta_d(\epsilon_d+p_d),
\end{equation}
\begin{equation}
\Gamma_u^2\beta^2_u(\epsilon_u+p_u)+p_u = \Gamma^2_d\beta^2_d
(\epsilon_d+p_d)+p_d.
\end{equation}
Number densities (n), pressure (p) and energy densities ($\epsilon$) are all measured in the comoving frame of 
the plasma we refer to, while the Lorentz factor ($\Gamma$) is measured in the shock frame. The indices 'u' 
and 'd' refer to the upstream and downstream plasmas respectively.

For simplicity, we consider the case of a strong shock (i.e., upstream plasma is cold), so that $p_u=0$ and 
$\epsilon\sim n_um$. We use an equation of state which was introduced by \citet{b18}. 

The basic assumption is that the plasma consists of a single component with  temperature $T$, 
\begin{equation}
\epsilon + p = \rho G \left(\frac{m}{k_BT} \right),
\end{equation}
where $G(x)=K_3(x)/K_2(x)$ and $K_2$,$K_3$ are the modified Bessel functions. If the downstream plasma can be 
well described as an ideal gas, Eq.~(4) can be rewritten as
\begin{equation}
\epsilon = \rho  G \left(\frac{nm}{p} \right) - p.
\end{equation}
Solving these equations, the solutions for $\beta_d$ are obtained. $\beta_u$ and $\beta_d$ are necessary for 
calculating $P_u$ and $P_d$.

\subsection{calculation of $P_u$ \& $P_d$}
In this subsection we briefly summarize the method of determining $P_u$ and $P_d$ in the case of both the 
large-angle scattering and the regular deflection. $P_u$ ($P_d$) is the conditional probability that a particle 
entering into the upstream (downstream) along a direction $\mu_0$ will leave toward the downstream (upstream) 
along a direction $\mu$. 

First, we explain about the case of the large-angle scattering which was proposed by \citet{b9}. They assumed 
the energy of a particle measured in the fluid frame is conserved by scattering. The probability of displacement 
$\Delta z$ of the particle along shock normal is defined as
\begin{equation}
p(\Delta z,v,\mu)d\Delta z=\frac{1}{\lambda}e^{-\frac{|\Delta z|}{\lambda}}d\Delta z.
\end{equation}
Here, $\lambda$ denotes a mean free path of the particle measured in the shock rest frame, $v$ is the speed of 
the particle and $\mu$ is the pitch angle cosine measured in the fluid frame. They use the scattering model that 
the pitch angle cosine of the particle $\mu$ is determined according to the probability density function $P_\mu(\mu,v)$. 
$P_\mu$ is independent of the initial pitch angle. This means the initial information is lost and there 
is no correlation between the initial distribution function and the last one. They consider the scattering is isotropic 
in the fluid frame:
\begin{equation}
P_\mu=\frac{1}{2}.
\end{equation}
Next, they defined the probability density function (p.d.f.) of the displacement for each step of the random walk as
\begin{eqnarray}
f(\Delta z) 
&=&\int_{-1}^{1}P_\mu p(\Delta z,v,\mu)d\mu \nonumber \\
&=&  \left\{\begin{array}{ll}
			\int_{-u}^{1} \frac{P_\mu}{\lambda} e^{-\frac{\Delta z}
                             {\lambda}}d\mu  \:(\Delta z > 0) &\quad  \\
			\int_{-1}^{-u} \frac{P_\mu}{\lambda} e^{\frac{\Delta z} 
                             {\lambda}}d\mu  \:(\Delta z < 0), &\quad
			\end{array}
			\right.
\end{eqnarray}
where $u$ is the fluid velocity (e.g., $u_d$ is the downstream fluid velocity in the shock rest frame). This p.d.f. 
expresses the probability of the particle which moves $\Delta z$ after the particle is scattered once. Then, we can 
calculate p.d.f. of scattering points at the \textit{m}-th step, $f_\textit{m} (\Delta z)$. Finally, they introduced 
the density of scattering points summed over all steps that is written as
\begin{equation}
n(\Delta z)=\sum_{\textit{m}=1}^{\infty} f_\textit{m}(\Delta z),
\end{equation}
where $n(\Delta z)$ means that the particle is translated by $\Delta z$ before it crosses the shock front. 
By using the formulation mentioned above, we can estimate $P_u$ and $P_d$ after some detailed calculations. The 
results are:
\begin{equation}
P_u(\mu_0,\mu)=C_0(\lambda_0)\frac{P_\mu \lambda}{\lambda + \lambda_0}+C_1
(\lambda_0)\frac{P_\mu \lambda}{\lambda + L_D},
\end{equation}
\begin{equation}
P_d(\mu_0,\mu)=C'_0(\lambda_0)\frac{P_\mu \lambda}{\lambda + \lambda_0} + C'_1
(\lambda_0) P_\mu \lambda,
\end{equation}
where $C_0,C_1,C'_0,C'_1$ are the functions which depend on the shock speed and the mean free path. $L_D$ is the 
diffusion length which is defined by \citet{b9}. The total return probability in the downstream, $P_R$, is not 
unity, and given as
\begin{equation}
P_R(\mu_0)=\int ^{-u}_{-1} P_d(\mu_0,\mu) d\mu.
\end{equation}
They also calculate the flux across the shock front. The results are 
\begin{equation}
\phi_{du}(\mu)=\frac{P_\mu \lambda}{h_-},
\end{equation}
\begin{equation}
\phi_{ud}(\mu)=\frac{1}{g_+(L_D)}\frac{P_\mu \lambda}{\lambda + L_D},
\end{equation}
(as for the estimation of $h_-$ and $g_+$, see \citet{b9}). $\phi_{du}$ ($\phi_{ud}$) is the flux entering to the 
upstream (downstream). 

Next, we explain about the case of the regular deflection in the upstream. The regular deflection is important for a 
relativistic shock in the upstream. In the case of the regular deflection, we have to solve the equation of motion 
in large-scale magnetic fields (\citet{b8} and \citet{b1}). Here, we choose the shock normal in the z-direction and 
assume the magnetic fields is parallel to the x-direction. We calculate the pitch-angles $\mu_1$ when particles cross 
the shock front from the upstream. First, we decide the initial azimuthal/pitch angles $\mu_0$ and $\phi_0$. The 
particle's velocity is decided as 
\begin{equation}
\beta_x=\cos \phi \sin \theta, 
\beta_y=\sin \phi \sin \theta, 
\beta_z=\cos \theta.
\end{equation}
Next, we solve the equation of motion which is written as 
\begin{equation}
\frac{d\textrm{\boldmath $\beta$}}{dt}=\Omega _g (\textrm{\boldmath $\beta$}\times 
\textrm{\boldmath $\hat{b}$}),
\end{equation}
where $q=Ze$ is charge of the particle, $E$ is energy and $\Omega_g=ZeBc/E$ is a gyration frequency in magnetic 
fields \textrm{\boldmath $B$}=B \textrm{\boldmath $\hat{b}$}. We solved this equation numerically using the above 
velocity as the initial condition. The upstream residence time $t_u$ is obtained by considering $z(t_u) = z_s = 
\beta_s t_u$. Here, $z(t_u)$ is the position of the particle and $z_s$ is the position of the shock front. Finally, 
we calculate the pitch-angle $\mu_1$ using Eq.~(15) again. $\mu_1$ is determined uniquely as a function of $\mu_0$. 
We can express $P_u$ using four angles as 
\begin{eqnarray}
P_u(\mu_0,\mu)
&=&(2\pi)^{-1}\delta(\mu-\mu_1(\mu_0,\phi_0)) \nonumber \\
& &\times \delta(\phi-\phi_1(\mu_0,\phi_0)).
\end{eqnarray} 

\subsection{Peacock's approximation}
\citet{b17} developed the calculating method to obtain a power-law index which was originally used by \citet{b4}. 
Peacock's approximation can be applied to a relativistic shock. He considered $N_0$ particles crossing from the 
upstream to the downstream with an initial energy $E_0$. We use a return probability $P_R(\mu_0)$ that a particle 
will eventually return to the upstream. The particle's energy is increased by a energy-gain factor $G(\mu_0,\mu)$, 
where $\mu_0$ and $\mu$ are the initial pitch angle and the last pitch angle for one step. After \textit{k} cycles, 
number of particles that remains in the upstream $(N)$ is expressed as
\begin{equation}
\frac{N}{N_0}=\langle P_R \rangle^k,
\end{equation}
where
\begin{equation}
\langle P_R \rangle = \int ^1_{-u_d}P_R(\mu_0)\phi_{ud}(\mu_0)d\mu_0.
\end{equation}
Here, $\langle P_R \rangle$ is the return probability averaged over $\mu_0$ with the weight of the flux crossing the 
shock with various values of $\mu_0$. As particles cross and re-cross the shock, the distribution of energies is 
broadened. Peacock's approximation assumes the numbers of cycles are large enough to use the central limit theorem. 
It is also necessary that the energy-gain of each step is fully uncorrelated. If particles cross the shock front many 
times, the distribution of energies can be expressed by a Gaussian by a central limit theorem. In such a situation, 
we can approximate particle's energy is amplified at the same rate per one cycle and the effect of the variance is 
neglected. We can calculate the power-law index only using the averaged energy-gain factor. After \textit{k} cycles, 
the particle energy is given by
\begin{equation}
\ln \left(\frac{E}{E_0} \right)=k\langle \ln G \rangle,
\end{equation}
where $G$ is the energy-gain factor and expressed as $G(\mu_{out},\mu)=\left(\frac{1-V_r\mu_{out}}{1-V_r\mu}\right)$. 
The averaged $\ln G$ is defined as:
\begin{eqnarray}
\langle \ln G \rangle 
&=& \int ^1_{-u_d}d\mu \int ^{-u_d}_{-1}d\mu_{out} \nonumber \\
& &\times \left[P_u(\mu_{out},\mu) \ln \left(\frac{1-V_r\mu_{out}}{1-V_r\mu}\right) \phi_{du}(\mu_{out})\right] \nonumber\\
& &\times \left[ \int ^{-u_d}_{-1}d\mu_{out}\phi_{du}(\mu_{out}) \right] ^{-1}.
\end{eqnarray}
Here, $V_r=\frac{u_u-u_d}{1-u_uu_d}$ indicates the relative velocity of the upstream fluid with respect to the downstream 
fluid. From Eq.~$(18)$ and $(20)$, the integrated energy spectrum is written as
\begin{equation}
\ln \left(\frac{N}{N_0} \right) = \frac{\ln \langle P_R \rangle}{\langle \ln G \rangle} \ln \left( \frac{E}{E_0} \right).
\end{equation}
Thus the differential energy spectrum is obtained as
\begin{equation}
f(E) d^3\textrm{\boldmath $p$} \propto E^{-s} d^3\textrm{\boldmath $p$},
\end{equation}
where
\begin{equation}
s=3-\frac{\ln \langle P_R \rangle}{\langle \ln G \rangle}.
\end{equation}

\subsection{Vietri's formulation}
Next, we explain derivation of a power-law spectrum given by \citet{b20}. He derived the relativistically covariant 
equation for the distribution function of particles accelerated at a shock, which is applicable to a relativistic 
shock. His formulation gives the exact power-law index and does not assume uncorrelation among various energy-gains. 
\citet{b6} solved the transport equation providing the boundary condition for the flux which crosses the shock front 
to the downstream ($\phi_{ud}$) 
(upstream ($\phi_{du}$)) by using particles which crosses the shock front to the upstream ($\phi_{du}$) (downstream 
($\phi_{ud}$)), and they derived $P_u$ and $P_d$. In this paper, we use $P_u$ and $P_d$ which are calculated in section 
2.2. The power-law index is given by this boundary condition:
\begin{eqnarray}
\phi_{ud}(\mu) 
&=&\int ^{-u_d}_{-1} d\mu_{out} P_u(\mu_{out},\mu) \nonumber \\
& &\times\left(\frac{1-V_r\mu}{1-V_r\mu_{out}}\right)^{3-s} \phi_{du}(\mu_{out}).
\end{eqnarray}

\begin{figure}
\centering
\includegraphics[height=7.5cm,angle=270,clip]{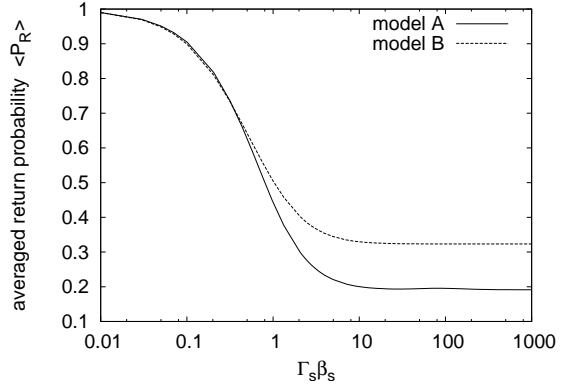}
\caption{Return probability vs the shock speed for model A (solid line) and model B (dotted line). 
We assume large-scale magnetic fields are present in the upstream, with a direction perpendicular 
to the shock normal when we consider the regular deflection (model B).} 
\end{figure}%
This equation can be integrated over the whole range of $\mu$ and divided by the whole flux entering into the 
upstream, which gives
\begin{eqnarray}
& & \frac{\int ^1_{-u_d}d\mu \phi_{ud}(\mu)}{\int ^{-u_d}_{-1}d\mu_{out} \phi_{du}(\mu_{out})} \nonumber \\
&=&  \int ^1_{-u_d}d\mu \int ^{-u_d}_{-1}d\mu_{out} \nonumber \\
& & \times \left[P_u(\mu_{out},\mu) \left(\frac{1-V_r\mu}{1-V_r\mu_{out}} \right)^{3-s}
\phi_{du}(\mu_{out})\right]  \nonumber \\
& & \times \left[ \int ^{-u_d}_{-1}d\mu_{out}\phi_{du}(\mu_{out}) \right] ^{-1}. 
\end{eqnarray} \\
The left term is the inverse of the averaged return probability from the downstream. The right term is the average of 
the $(s-3)$ power of the energy-gain factor $G$. Eq.~$(26)$ can be rewritten as
\begin{equation}
\langle P_R \rangle \langle G^{s-3} \rangle=1.
\end{equation}

In Newtonian limit, we can approximate $ \langle G^{s-3} \rangle \sim \langle G \rangle ^{s-3}$ because $G-1\ll1$ 
for a non-relativistic shock. Thus the energy spectral index is given by the similar form as $Eq.~(24)$. Vietri 
insisted that $ \langle G^{s-3} \rangle $ can not be approximated by $\langle G \rangle ^{s-3}$ in general for a 
relativistic shock, contrary to the Peacock's argument. 

\section{Results}
In this section we calculate a power-law index using Peacock's approximation and Vietri's formulation. Then we 
examine the characteristic of the power-law index and how the power-law index derived from Peacock's approximation 
is different from the one derived from Vietri's method to examine the effect of the variance. We consider model A 
(where the large-angle scattering occurs in both the upstream and the downstream), and model B (where the large-angle 
scattering occurs in the downstream and particles are deflected by large-scale magnetic fields in the upstream). 
We need the averaged chance probability for a particle to return from 
\begin{figure}
\centering
\includegraphics[height=7.5cm,angle=270,clip]{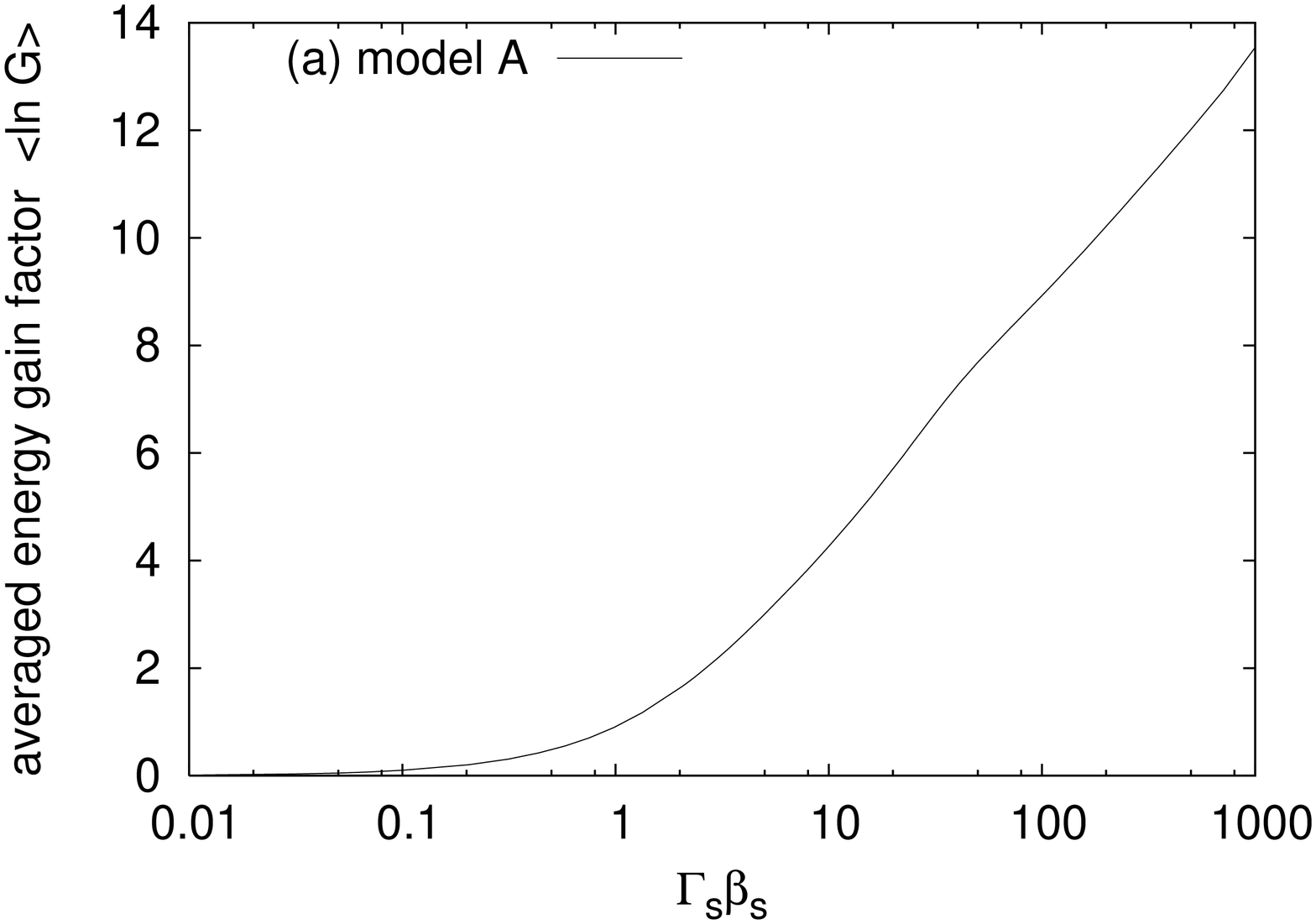}
\includegraphics[height=7.5cm,angle=270,clip]{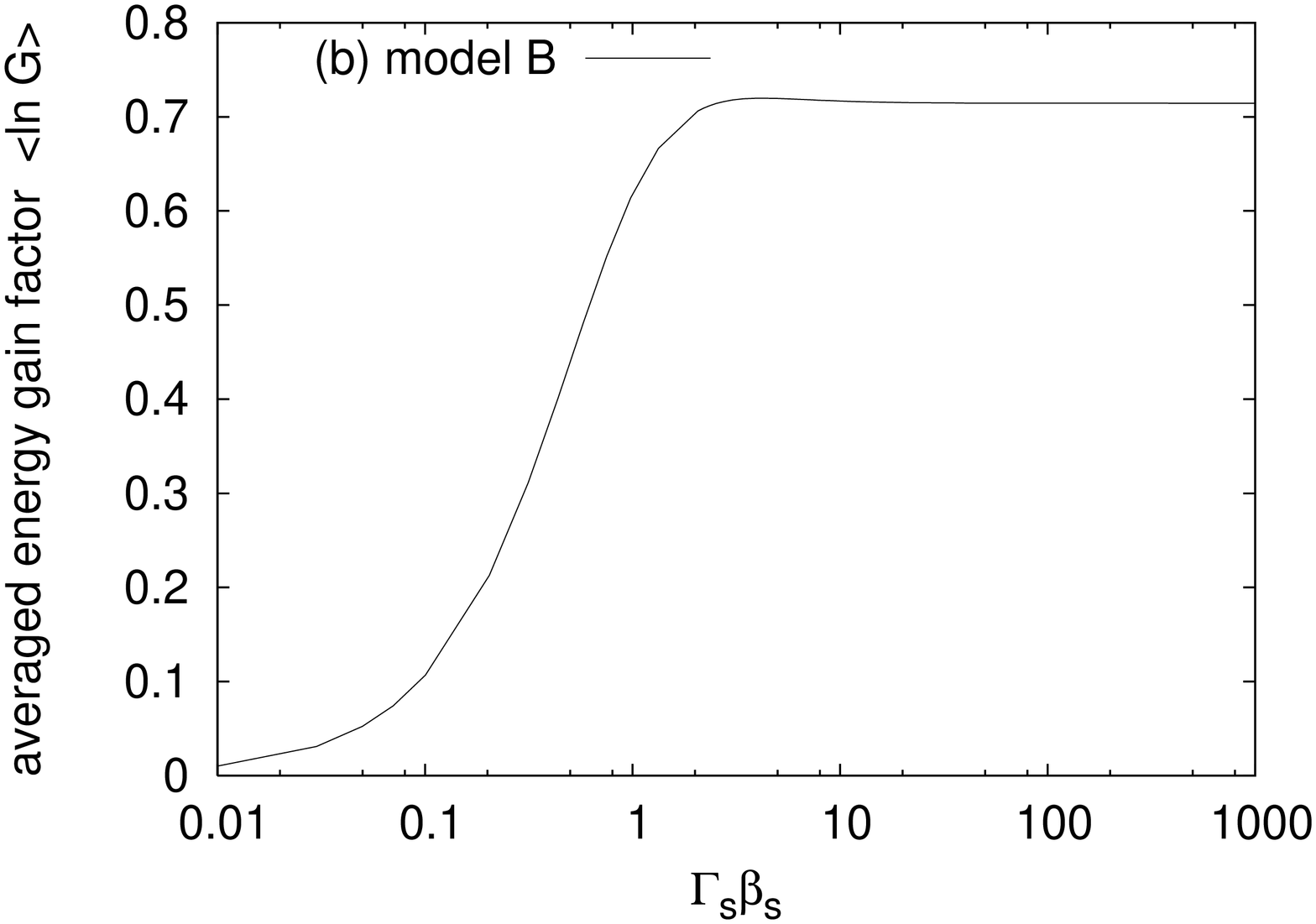}
\caption{The averaged energy-gain factor for model A (top panel) 
and model B (bottom panel). We can see the averaged energy-gain factor of model A 
is much larger than that of model B.}
\end{figure}
the downstream to the upstream per a crossing cycle 
$\langle P_R \rangle$ (Eq.~(19)) and the averaged energy-gain factor $\langle \ln G \rangle$ 
(Eq.~(21)) to calculate the power-law index by Peacock's approximation. 
In Fig.~1 we show the averaged return probability. In the both cases 
of model A and model B, the return probability converges along the shock velocity. This is because the shock velocity 
converges to one third of the speed of light in the down stream when we use the Synge equation as the equation of the 
state. So the probability for particles to catch up with the shock front converges. In Fig.~2 we show the averaged 
energy gain factor. The value increases monotonically in model A (Fig.~2 top). On the other hand, the value converges 
in model B (Fig.~2 bottom). Particles get more energy by the shock-acceleration when the shock velocity becomes faster and the scattering angle becomes 
larger. In model A, particles are scattered at a large angle and get large energy as the shock moves with the 
large velocity. In model B, the deflection angle becomes smaller as the shock velocity becomes large. Both of the two 
effects (a large velocity and a small scattering angle) cancel out each other and the energy amplification converges. 
In Fig.~3, we show the power-law index. The power-law index becomes harder as a shock moves fast in model A. This is 
because the averaged energy-gain factor increases monotonically. On the other hand, the power-law index converges in 
model B. This is due to the convergence of the averaged return probability and the averaged energy-gain factor. This 
convergence is the same behavior as \citet{b3}. They studied about the case of the 
\begin{figure}
\centering
\includegraphics[height=7.5cm,angle=270,clip]{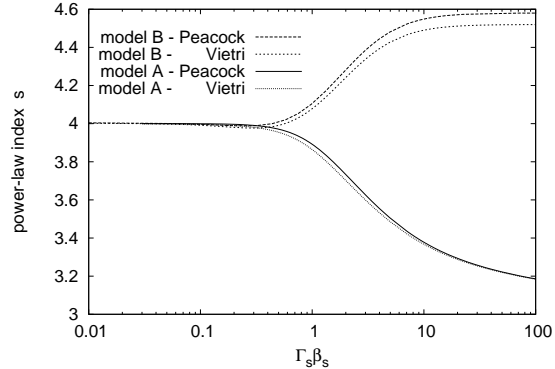}
\caption{The power-law index vs. the shock speed. The power-law indices in model A and model B 
are calculated by Peacock's approximation and Vietri's formulation. The above two lines correspond to model B, 
while bottom two lines correspond to model A.}
\end{figure}
small-angle scattering and the 
power-law index converges as the shock velocity becomes large. The power-law index 
converges to 4 for a non-relativistic 
case which is consistent with the result of the diffusive shock-acceleration. 

Next, we consider the difference between the power-law index derived from Peacock's approximation and Vietri's 
formulation. The power-law index converges in model B, but the converged values are different between Peacock's 
approximation and Vietri's formulation. So Peacock's approximation is not suitable for relativistic shocks. On the 
other hand, in the case of model A, the difference of the power-law index is the largest when $\Gamma_s \beta_s$ 
is about 3, and the index becomes nearly equal when $\Gamma_s \beta_s$ is greater than 10. We can use Peacock's 
approximation for highly-relativistic shocks. We also show that the power-law index calculated by Vietri's 
formulation is harder than the one calculated by Peacock's approximation in both cases of model A and model B. 
We explain why the power-law spectrum becomes hard in the discussion. 

Next we see what makes the difference between Peacock's approximation and Vietri's formulation.First, transform 
Eq.~(24) and Eq.~(27) as 
\begin{equation}
\ln \langle P_r(\beta_{s_1}) \rangle+\ln \langle G^{s_1-3}(\beta_{s_1}) \rangle = 0,
\end{equation}
\begin{equation}
\ln \langle P_r(\beta_{s_1}) \rangle + (s_2-3) \langle \ln G(\beta_{s_1}) \rangle = 0,
\end{equation}
where $s_1(s_2)$ is the power-law index derived from Vietri's formulation (Peacock's approximation) and $\beta_{s_1}$ 
is the shock velocity which satisfies Eq.~(27). Peacock's approximation (Eq.~(24)) and Vietri's formulation (Eq.~(27)) 
coincide with each other as long as the following equation is satisfied: 
\begin{figure}
\centering
\includegraphics[height=6.5cm,angle=270,clip]{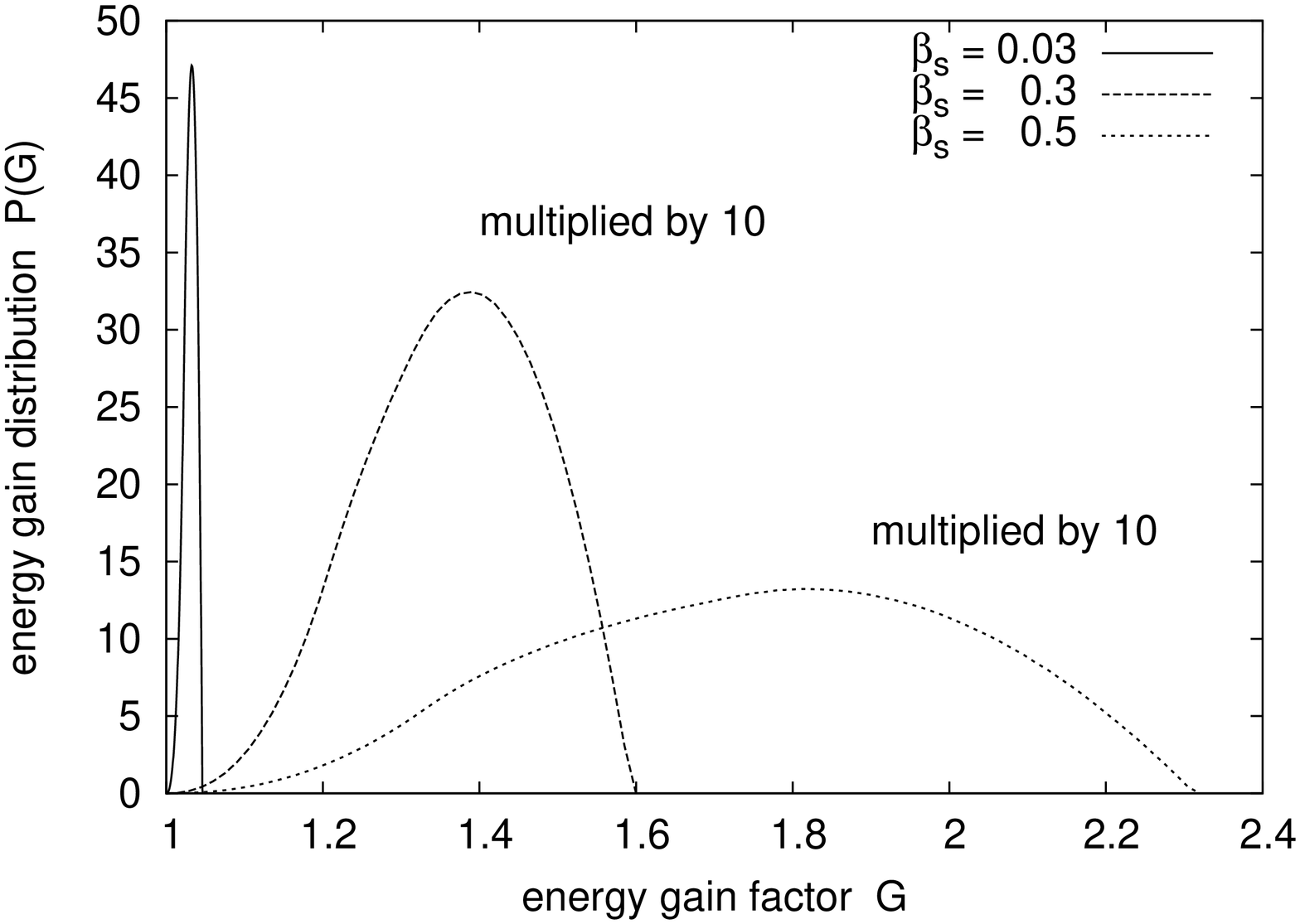}
\includegraphics[height=6.5cm,angle=270,clip]{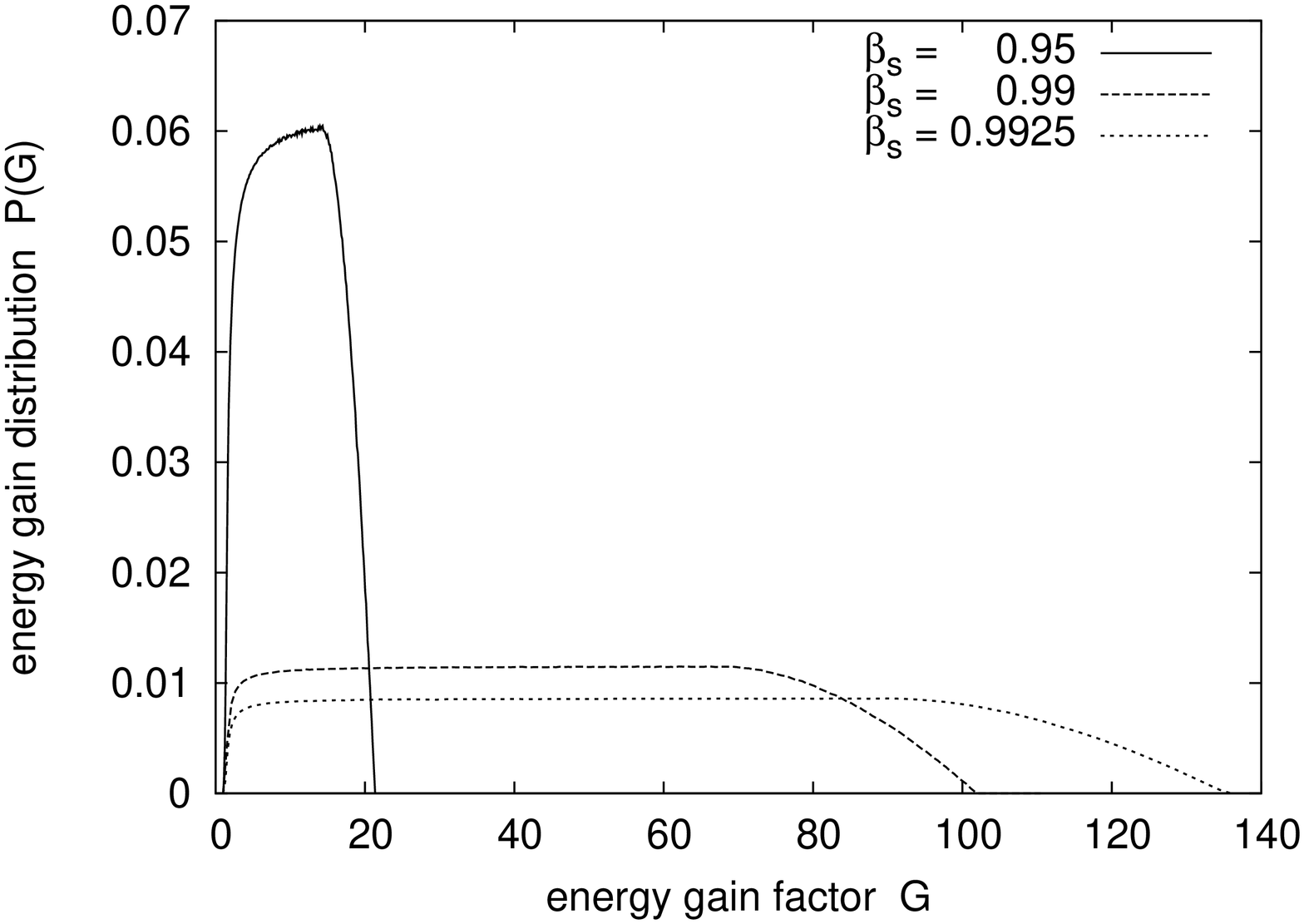}
\includegraphics[height=6.5cm,angle=270,clip]{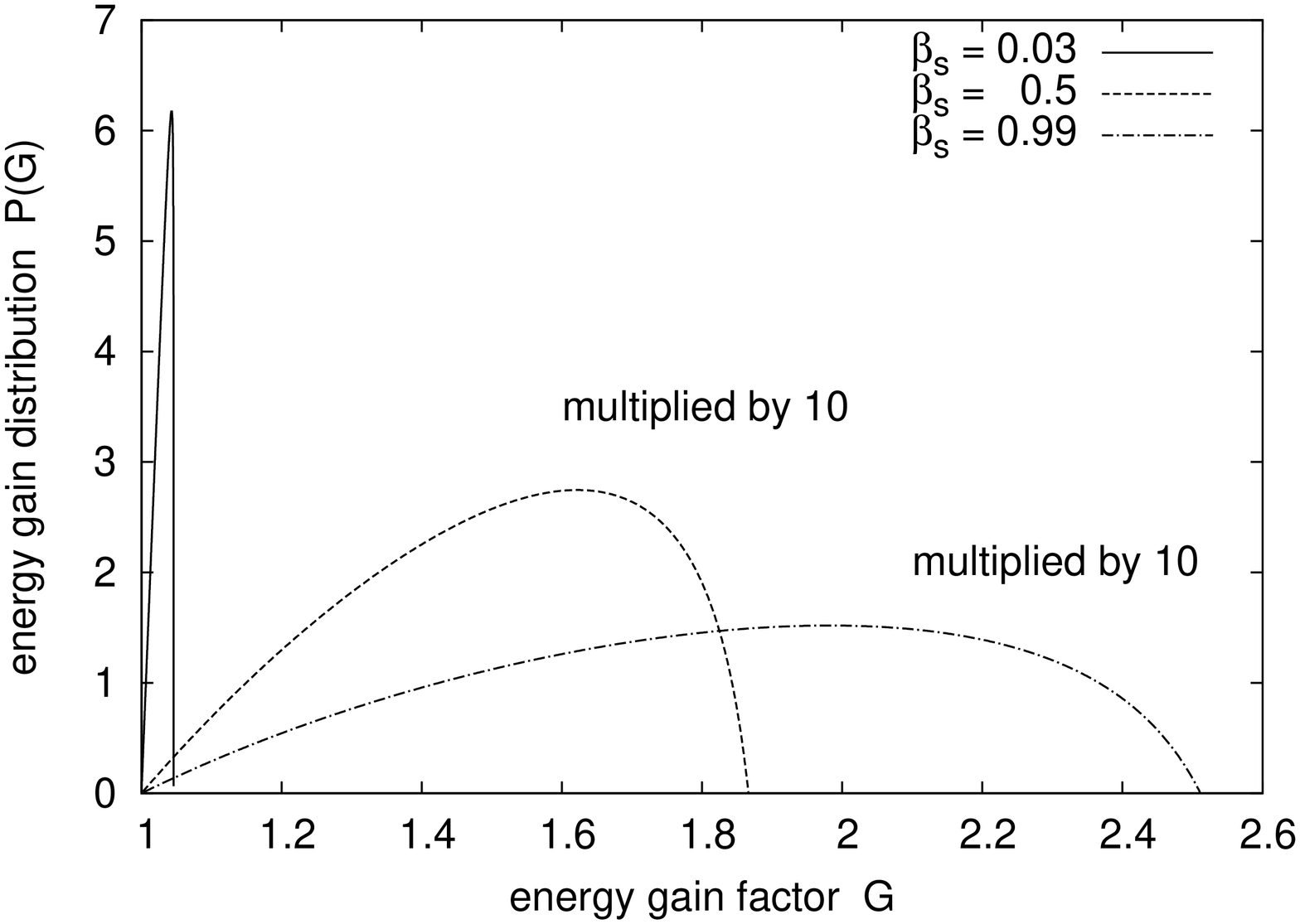}
\caption{The energy-gain factor's distribution function $P(G)$ in 
model A (top panel and middle panel) and model B (bottom panel). 
The results are plotted in the cases of several values of the shock velocity. In model A, the plotted distribution 
function is rescaled (multiplied by 10) both for $\beta_s=0.3$ and $\beta_s=0.5$. At the low velocity (top panel), 
the distribution function does not looks like a rectangle. At the high velocity (middle panel), the distribution 
function looks like a rectangle and we use the rectangular distribution function. In model B, we plotted the 
distribution function when $\phi$ is $\frac{3}{2}\pi$. The plotted distribution is rescaled (multiplied by 10) 
both for $\beta_s=0.5$ and $\beta_s=0.99$. In this case, the distribution does not look like a rectangle and we 
can not use approximation.}
\end{figure}
\begin{figure}
\centering
\includegraphics[height=7.5cm,angle=270,clip]{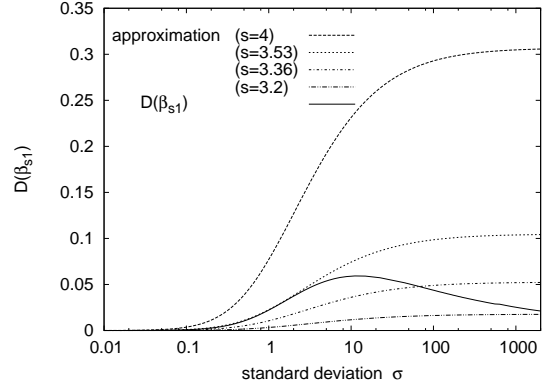}
\caption{$D$ and $D_{app}$ vs. the standard deviation. 
The approximation gives the contour with the power-law index fixed. 
It could be understood that the effect of the variance 
becomes weak as the power-law index approaches 3.}
\end{figure}
\begin{equation}
\ln \langle G^{s_1-3}(\beta_{s_1}) \rangle =(s_1-3)\langle \ln G (\beta_{s_1})\rangle.
\end{equation}
Now we introduce the value defined as 
\begin{equation}
D(\beta_{s_1}) = \ln \langle G^{s_1-3}(\beta_{s_1}) \rangle -(s_1-3)\langle \ln G(\beta_{s1}) \rangle .
\end{equation}
We use $D(\beta_{s_1})$ as the function which indicates the difference between Peacock's approximation and Vietri's 
formulation. The averaged value is calculated using the energy-gain factor distribution as follows,
\begin{equation}
\langle G \rangle = \int G P(G) dG,
\end{equation}
where
\begin{equation}
P(G)=\int ^1_{-u_d}d\mu \left(\frac{1-V_r\mu}{V_r}\right)P_u\phi_{du}. 
\end{equation}
$D$ becomes $0$ if Peacock's approximation and Vietri's formulation give the same result. Particularly, $D$ 
approaches 0 when $s$ approaches 3. $D$ becomes large if the difference between Peacock's approximation and 
Vietri's formulation becomes conspicuous. 

Now let us see the effect of the variance in $D$. However, it is difficult to know the relation between $D$ and 
the variance. 
Here, we use the simple model of the distribution function which shows the distribution function (Eq. (33)) 
approximately. In Fig.~4 the energy-gain factor's distribution function $P(G)$ is plotted for the various shock 
velocity. In the case of model A, the distribution becomes wider and rectangular as the shock moves fast. Then 
we use the rectangular distribution function for highly-relativistic shocks in model A. The rectangular distribution 
function is written as, 
\begin{equation}
P_{app}(G) = 
\begin{cases}
\frac{1}{L-1}& \text{$1 \le G \le L$} \\
0& \text{other.}	
\end{cases}	
\end{equation}
On the other hand, the distribution does not become rectangular and seems unsuitable to apply this simple distribution 
function in model B. $D$, which is calculated using the simple distribution model, is written as, 
\begin{eqnarray}
D_{app}(L,s_1) &=&\ln\frac{1}{s_1-2}\left[\frac{L^{s_1-2}}{L-1}-\frac{1}{L-1}\right] \nonumber \\
               & &-\frac{s_1-3}{L-1}\left(L\ln L - L +1 \right). 
\end{eqnarray}
We expect that the variance which is calculated by the true distribution function (Eq.~(33)) corresponds to the one 
which is calculated by the rectangular distribution. L has a relation to the variance $\sigma^2$ as 
$\sigma=(L-1)/2\sqrt{3}$. Using this relation, $D_{app}(L,s_1)$ is rewritten as,
\begin{eqnarray}
& &D_{app}(\sigma,s_1) \nonumber \\
&=&\ln\frac{1}{s_1-2}\left[\frac{(2\sqrt{3}\sigma+1)^{s_1-2}}{2\sqrt{3}\sigma}-
\frac{1}{2\sqrt{3}\sigma}\right] \nonumber \\
& &-\frac{s_1-3}{2\sqrt{3}\sigma}\left((2\sqrt{3}\sigma+1)\ln (2\sqrt{3}\sigma+1) 
- 2\sqrt{3}\sigma \right).
\end{eqnarray}
$D_{app}$ has the convergence value when $\sigma$ becomes infinity. The convergence value is written as, 
\begin{equation}
D_{app}(\infty,s_1)=\ln\frac{1}{s_1-2}+s_1-3.
\end{equation} 

We examine how D depends on the variance in model A. In Fig.~5, we can plot $D$ and $D_{app}$ as the functions of 
the standard deviation (i.e., the variance) instead of $\beta_{s1}$ because the variable increases monotonically 
as $\beta_{s1}$ increases. We show the relation between the shock velocity and the variance in Fig.~6. There is 
a one-to-one correspondence 
\begin{figure}
\centering
\includegraphics[height=7.5cm,angle=270,clip]{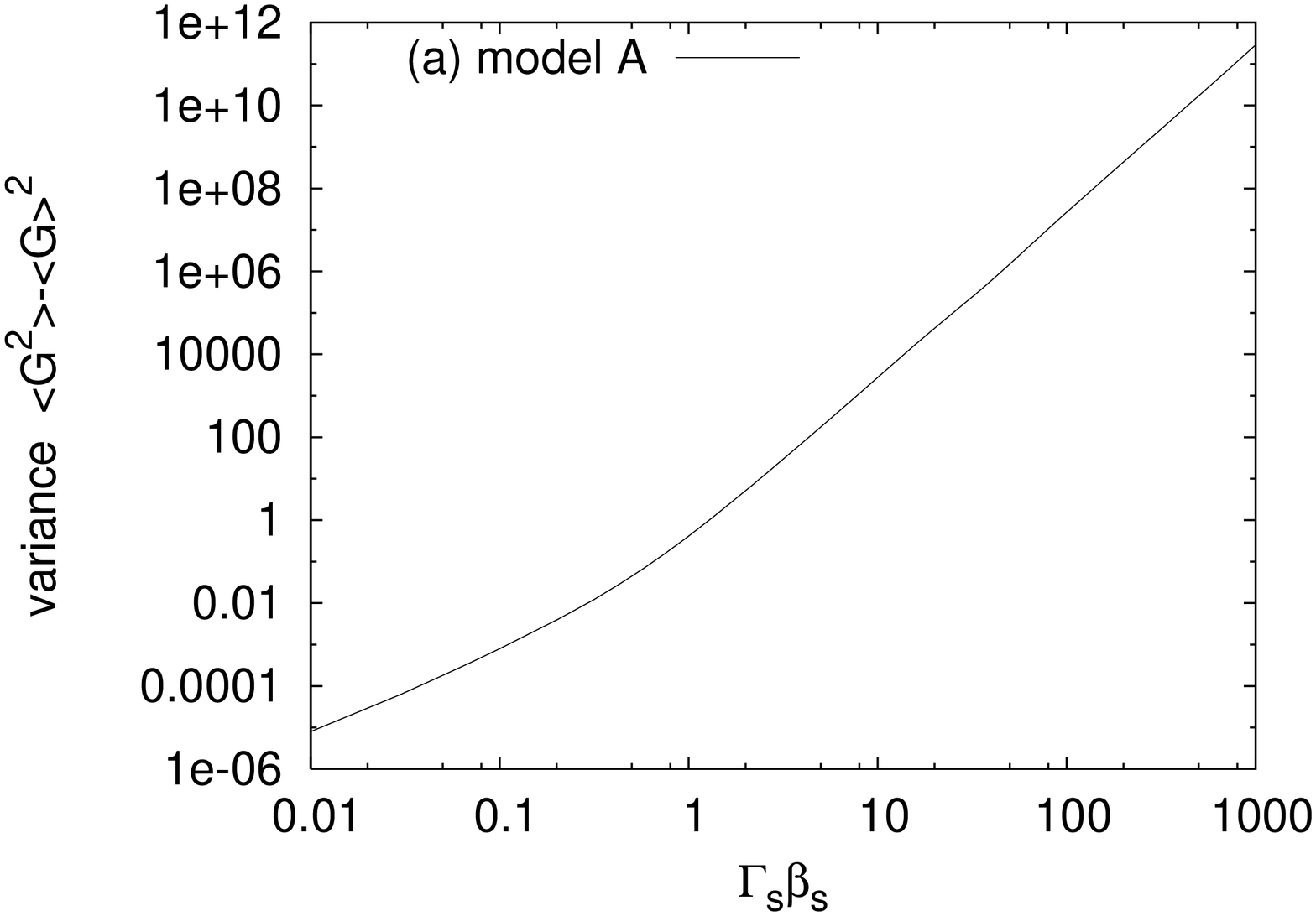}
\includegraphics[height=7.5cm,angle=270,clip]{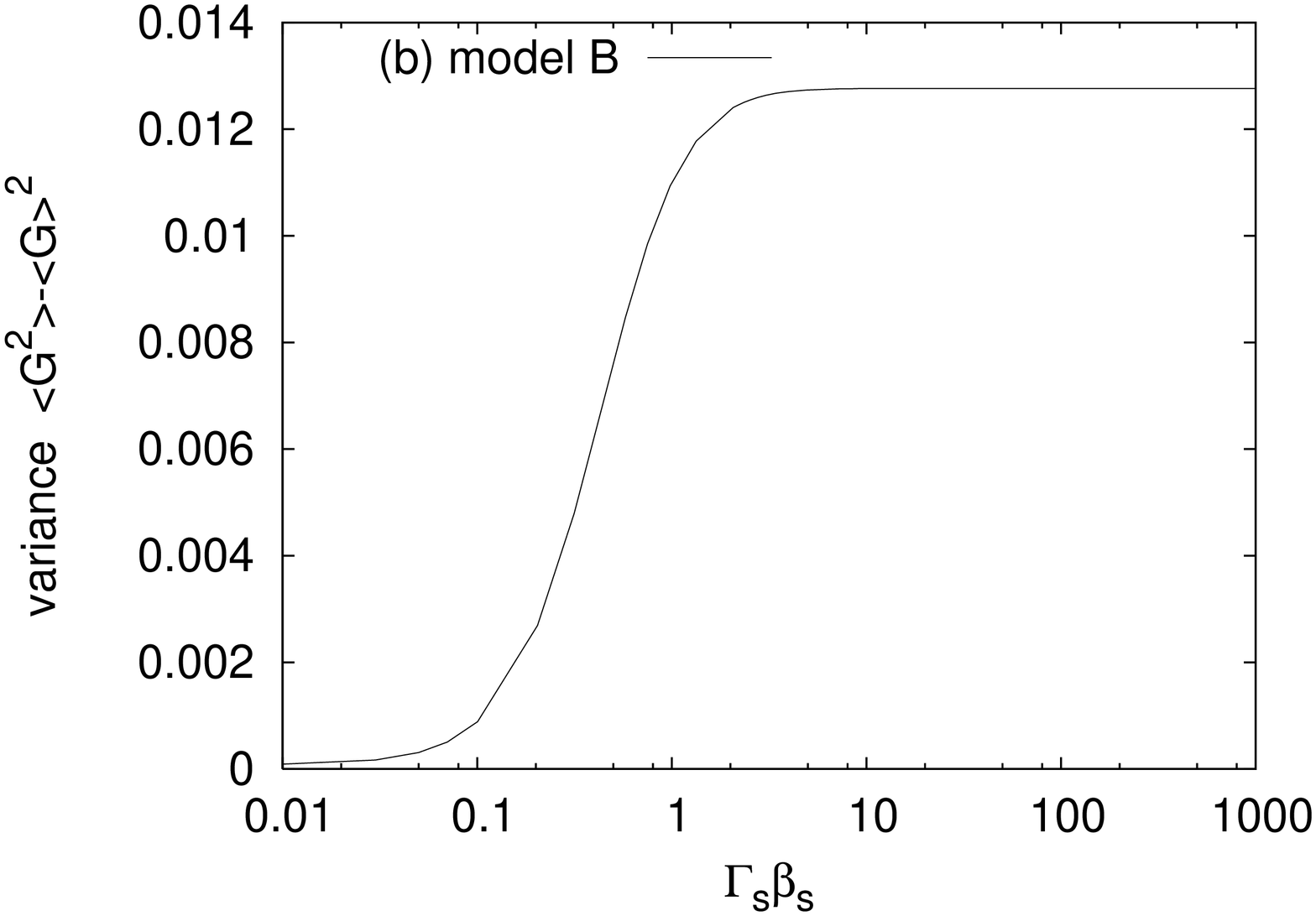}
\caption{The variance of the energy-gain factor's distribution function vs. the shock speed for model A (top panel) 
and model B (bottom panel).}
\end{figure}
between the variance and the shock speed. $\sigma$ and $s$ is written as the functions 
of one variable as, $\sigma=\sigma(\Gamma\beta)$, $s=s(\Gamma\beta)=s(\sigma)$. We calculate $D$ using $s_1$ which 
is calculated by Eq.~(27). We plot the contour of $D_{app}$ fixing $s_1$ mathematically in order to examine the 
effect of the variance. $D_{app}$ decreases as $s_1$ approaches 3 for the fixed variance. This means the effect 
of the variance becomes weak as $s_1$ approaches 3. $D_{app}$ gives inadequate approximation when the shock speed 
is small. In fact, $D_{app}$ caluculated by $s_1=3.53$ gives the same 
value as $D$ although $D_{app}$ calculated by $s_1=4$ should give the same value as $D$. For highly-ralativistic, 
shocks, the approximation becomes good. $D$ changes along the contour which is derived from the rectangular approximation 
at first because the power-law index does not change when the shock speed is low. $D$ crosses the contour as the shock 
moves fast because the power-law index approaches 3. This behavior shows that the effect of the variance becomes weak 
by the approach of $s$ to 3. 

In Fig.~7, $D$ is plotted both in model A and model B as the functions of $\Gamma_s\beta_s$. In model A, $D$ becomes 
the largest when $\Gamma_s\beta_s$ is about 3, and the variance is effective. Although the variance increases 
monotonically, $D$ decreases for highly-relativistic shocks. This is because the effect of the variance becomes weak 
as the power-law index approaches 3. $D$ becomes 0 for non-relativistic shocks because the variance becomes small 
as we can see from Fig.~6. In model B, we can interpret behavior of $D$ as explained before with respect to the 
variance and the power-law index. $D$ increases as the variance increases at first because the variance increases 
\begin{figure}
\centering
\includegraphics[height=7.5cm,angle=270,clip]{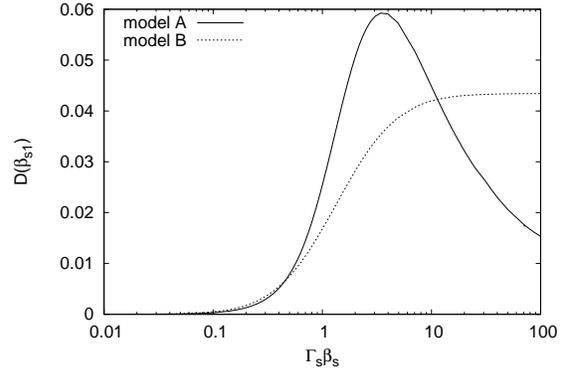}
\caption{$D$ is plotted both in model A and model B. We can interpret that $D$ is derived from the effect of the 
variance in model B. In model A, $D$ has maximum value . In model B, $D$ converges as the shock speed becomes large.}
\end{figure}
as the shock velocity becomes large (see Fig.~6). After the variance converges, 
$D$ continues to increase until $\Gamma_s\beta_s$ becomes as high as 30. This is because the power-law index 
increases and the effect of the variance becomes strong until $\Gamma_s\beta_s$ becomes as high as 30. For 
non-relativistic shocks, $D$ also becomes 0 in model B.

\citet{b10} considered the effect of the variance in their shock-acceleration formulation. The effect of the variance 
also becomes negligible for a non-relativistic shock in the case of their formulation.

\section{Discussion \& Conclusion}

In this paper, we examine how the power-law index derived from Peacock's approximation is different from the one 
derived from Vietri's formulation. \citet{b17} gives the approximate power-law index and \citet{b20} gives the exact 
power-law index. \citet{b20} claimed that Peacock's approximation and Vietri's formulation give the different result 
in general, but they did not explain the reason in detail (see also \citet{b6}). The effect of the variance has not 
been studied well until now and we have examined the effect of the variance.

First, we conclude the difference of the power-law index derived from Peacock's approximation and Vietri's formulation. 
We considered two cases for the scattering in the upstream to model multiple shock crossing: the large-angle scattering 
(model A) and the regular deflection by large-scale magnetic fields (model B). We consider only the large-angle 
scattering in the downstream. In model A, the power-law index derived from Peacock's approximation is different 
from the one derived from Vietri's formulation for a mildly-relativistic shock. The difference decreases as the 
shock velocity approaches the highly-relativistic velocity. In model B, the difference becomes larger along the shock 
velocity at first. Finally, the difference converges when the shock velocity is mildly-relativistic. From the above 
result, we can use Peacock's approximation for a highly-relativistic shock when we consider model A and the difference 
between Peacock's approximation and Vietri's formulation is important when we consider model B for a relativistic shock. 
Moreover, we see the reason why Peacock's approximation and Vietri's formulation give the different results. We conclude 
the variance in the distribution of the energy gain factor affects the difference between Peacock's approximation and 
Vietri's formulation in model A. 

In Fig.~3, we show that the power-law index calculated by Vietri's formulation is harder than the one calculated by 
Peacock's approximation in both cases of model A and model B. This tendency can be understood as follows: Let us 
consider the number of particles with $E_k$, which have experienced $k$ cycles in an average (see Eq.~(20)). If there 
is the variance of energy-gain factor's distribution, there should be contribution to the number from those that have 
experienced less than $k$-cycles and more than $k$-cycles. It is apparent that the contribution from the former is 
larger than the latter, because the energy spectrum obeys a power-law with index larger than 3. As a result, the spectrum 
becomes hard when the effect of the distribution width is taken into account. As above, we can understand why the 
power-law index becomes hard as the variance becomes large. However, this interpretation can not be suitable when the 
variance is larger than some value. The effect of the variance converges as we show in Fig.~5. What's more, we can 
interpret that the effect of the distribution's variance becomes weak as the power-law index approaches 3 because the 
fraction of particles that are accelerated by the mean energy-gain factor increases. 

Next, let us explain the general characteristics of the power-law index when we change the scattering model. We found 
the power-law index converges as the shock velocity increases in model B. This is the same characteristic as 
\citet{b3} who studied about the case of the small-angle scattering, although the spectrum derived from the large-angle 
scattering is steeper than that of the small-angle scattering. On the other hand, the index decreases monotonically 
in model A. For the limit of non-relativistic shocks, the result is consistent with the diffusive shock-acceleration 
theory. 

The difference of the power-law index between model A and model B is large for a relativistic shock. The difference is 
small if one uses the small-angle scattering model instead of the large-angle scattering (e.g., \citet{b1}). This is 
because the averaged energy-gain factor is very large in model A. The energy-gain factor can be written as 
$G(\mu_{out},\mu)= \left( \frac{1-V_r\mu_{out}}{1-V_r\mu} \right)$. Thus, the energy-gain factor can be large for a 
highly-relativistic shock in the case of the large-angle scattering since the difference between $\mu_{out}$ and 
$\mu$ is large in the case of the large-angle scattering. On the other hand, the energy-gain factor remains the order 
of unity even for a highly-relativistic shock both in the case of the small-angle scattering and the regular deflection.

Peacock's approximation and Vietri's formulation give the different results in general. To see how Peacock's 
approximation is good, of course it is also important to investigate how many times scatterings should occur enough 
to use the central limit theorem and whether there is correlation among energy-gains for every step. Such consideration 
about these effects should be done not only for relativistic shocks but also for non-relativistic shocks. For example, 
we probably can not use a central limit theorem even for non-relativistic shocks in low energy range.This consideration 
might be useful to discuss the boundary between the supra-thermal particle and the non-thermal particle. The variance 
might also be considered to calculate the power-law index as we show in this paper.

We have considered the case of the large-angle scattering in this paper. We will consider the case of the small-angle 
scattering in the future work. 

\section*{Acknowledgement}
We thank Dr. T. Kato for valuable comments and discussion. 
The numerical calculations were carried out on 
Compaq Alpha Server ES40 at Yukawa Institute for
Theoretical Physics, Kyoto University. 
This work is in part supported by a Grant-in-Aid for the 21st Century
COE ``Center for Diversity and Universality in Physics'' from the
Ministry of Education, Culture, Sports, Science and Technology of
Japan.

\bsp

\label{lastpage}

\end{document}